\documentclass[aps,prc,twocolumn,showpacs,superscriptaddress,amsmath,amssymb]{revtex4}

\usepackage{CJK}
\usepackage{graphicx}
\usepackage{dcolumn}
\usepackage{bm}
\usepackage{hyperref}
\usepackage{multirow}
\usepackage{rotating}
\usepackage{url}
\usepackage{color}
\usepackage{ulem}

\newcommand{\delete}[1]{\bgroup\markoverwith{\textcolor{red}{\rule[0.5ex]{2pt}{1pt}}}\ULon{#1}}

\allowdisplaybreaks

\begin{document}

\title{Critical parameters of liquid-gas phase transition in thermal symmetric and asymmetric nuclear matter}

\author{Shen Yang}
\affiliation{School of Nuclear Science and Technology, Lanzhou University, Lanzhou 730000, China}
\affiliation{Institute of Modern Physics, Chinese Academy of Sciences, Lanzhou 730000, China}
\author{Bo Nan Zhang}
\affiliation{School of Nuclear Science and Technology, Lanzhou University, Lanzhou 730000, China}
\author{Bao Yuan Sun\footnote{sunby@lzu.edu.cn}}
\affiliation
{School of Nuclear Science and Technology, Lanzhou University, Lanzhou 730000, China}

\date{\today}

\begin{abstract}
The properties of critical parameters and phase diagram structure of liquid-gas phase transition are investigated in thermal symmetric and asymmetric nuclear matter with the covariant density functional (CDF) theory. Although uncertainty remains in predicting the critical parameters such as the critical temperature and pressure from various CDF functionals, several correlations are explored numerically and verified to be approximately linear between them. These correlations become worse when nuclear matter is more isospin asymmetric, resulting mainly from the effects induced by symmetry energy. By looking over the isospin dependence of the critical temperature, the role of the symmetry energy in LG transition properties of asymmetric matter is realized. The change of critical temperature with isospin asymmetry is found to be correlated well with and as a consequence could be constrained by the density slope of symmetry energy at saturation density. Then, the structure of phase diagram of thermal nuclear matter is analyzed carefully. It is revealed that the contribution from symmetry energy dominates the size of liquid-gas phase coexistence area. Moreover, the specific pattern of the phase diagram could be determined by the critical temperature at non-zero isospin asymmetry, illustrated from the correlations of the temperature with pressures at several characteristic points, paving the possible way to further explore the structure of liquid-gas phase diagram of thermal nuclear matter.
\end{abstract}


\maketitle

\section{Introduction}
Features of nuclear matter at finite temperature are of fundamental importance in nuclear physics. Among them, the so-called liquid-gas (LG) phase transition in sub-saturated thermal nuclear matter has drew a lot of attention, which  occurs due to the van der Waals behavior of the nucleon-nucleon interaction \cite{Siemens1983nature}. The LG phase transition of thermal nuclear matter have been studied both experimentally and theoretically in a variety of works over the past several decades\cite{Finn1982PRL, Siemens1983nature, Panagiotou1984PRL, Jaqaman1984PRC, Kapusta1984PRC, SRuKeng1987PRC, Muller1995PRC, Baldo1999PRC, MaYG1999PRL, Natowitz2002PRL, Gobet2002PRL, Karnaukhov2003PRC, Chomaz2004PhysRep, Das2005PhysRep, Pichon2006NPA, Brown2007PhysRep, Li2008PhysRep, Elliott2013PRC, BORDERIE2019PPNP}, and its important impact has been illustrated on many aspects of nuclear physics, such as heavy ion collisions\cite{Chomaz2004PhysRep, Das2005PhysRep, Brown2007PhysRep, Li2008PhysRep} and nuclear astrophysics\cite{Pethick1992RMP, PRAKASH1997Rep,Lattimer2004SCIENCE, Lattimer2016REP}.

The occurrence of LG phase transitions has been confirmed in both symmetric and asymmetric nuclear matter, being recognized from the survey of nuclear caloric curve and multifragment distribution in heavy-ion collisions \cite{Finn1982PRL, Panagiotou1984PRL, Natowitz2002PRL, Gobet2002PRL, Fevre2005PRL, Sfienti2009PRL, Elliott2013PRC}.
In the works for symmetric nuclear matter, one usually concerned the critical temperature $T_C$ of LG phase transition as an important and characteristic quantity \cite{SRuKeng1987PRC, Baldo1999PRC, Rios2008PRC, Rios2010NPA, Lourenco2016PRC}.
In general, $T_C$ is predicted with a large uncertainty, around $10\thicksim20$~MeV, from several theoretical models of thermal nuclear matter\cite{SRuKeng1987PRC, Muller1995PRC, ZuoW2004PRC, Rios2008PRC, JunXu2007PLB, Rios2010NPA, Vovchenko2015PRC, Vovchenko2017PRC}. On the other hand, its precise value is hardly constrained experimentally as well \cite{Panagiotou1984PRL, Natowitz2002PRL, Karnaukhov2003PRC, Elliott2013PRC}.
First, the uncertainty comes from the limitation that the available experiments only extrapolate $T_C$ to infinite matter from fragmentation reactions on finite nuclei \cite{Natowitz2002PRL, Das2005PhysRep, Sfienti2009PRL}.
Moreover, the finite-size effects are also found to be influential in estimating the critical parameters for experiments, where a model dependence is involved inevitably \cite{Moretto2003PRC, Moretto2005PRL, Elliott2013PRC}. The knowledge about the dependence of $T_C$ on various bulk properties of nuclear matter then would make sense on eliminating such an uncertainty.
It was revealed that $T_C$ could be correlated with the incompressibility of symmetric nuclear matter at zero temperature and the nucleon effective mass at nuclear saturation density $\rho_0$, paving the way to deduce $T_C$ by constrain related quantities \cite{Lourenco2016PRC}. Recently, one also concerned the correlation among critical parameters themselves, i.e., critical temperature $T_C$, critical density $\rho_C$ and critical pressure $P_C$, to understand the behaviors of LG phase transition \cite{Rios2010NPA, Elliott2013PRC}.

Recently, a lot of experimental efforts have been made on nuclear LG phase transition with extreme interest in its isospin asymmetry dependence \cite{Fevre2005PRL, Sfienti2009PRL, MCINTOSH2013PLB, McIntosh2013PRC}. With various kinetic thermometer approaches, the dependence of the nuclear caloric curve on the neutron-proton asymmetry can be extracted, correspondingly providing experimental information on the limiting temperatures of finite nuclei which is correlated with the critical temperature of nuclear matter \cite{Natowitz2002PRL}. For asymmetric nuclear matter, various theoretical works predict a phase diagram structure of LG phase transition in a wide range of isospin asymmetry and pressure \cite{Muller1995PRC, JunXu2007PLB, Dutra2008PRC, Sharma2010PRC, JWeiZhou2013PLB, Lenske2015PRC}. Some of them argued that the phase diagram properties of LG phase transition could be correlated and affected by the bulk properties of nuclear matter, especially the symmetry energy \cite{Muller1995PRC, JunXu2007PLB, Sharma2010PRC, JWeiZhou2013PLB}. However, the quantitative evidence is still not ample to deduce a clear correlation among each characteristic quantities of the LG phase diagram.

To investigate the LG phase transition of thermal nuclear matter, the covariant density functional (CDF) theory \cite{Serot1986ANP, Reinhard1989PPP, Ring1996PPNP, Bender2003RMP, Vretenar2005Rep, Meng2006PPNP, Niksic2006PPNP, Meng2015JPG, Meng2016} has been extended to the case of finite temperature, with many important achievements in describing the EoS of thermal nuclear matter\cite{Waldhauser1987PRC, Waldhauser1988PRC, Zimanyi1990PRC, Delfino1995PRC, Avancini2004PRC, Typel2010PRC, ShenG2010PRC}, the physics of supernova and proto-neutron star\cite{ShenH1998PTP, ShenH1998NPA, Broderick2000APJ, Menezes2003PRC, Avancini2008PRC, Avancini2010PRC, ShenG2011PRC, Hempel2012APJ}, and properties of excited hot nuclei\cite{Gambhir2000PRC, NiuYF2009PLB, NiuYF2011PRC, NiuYF2013PRC, LiuLang2015PRC, ZhangW2017PRC, ZhangW2018PRC}, etc..
The critical temperatures of LG phase transition within the CDF calculations, in general locating around $T_C\thickapprox15$~MeV, still keep model dependence \cite{Muller1995PRC, Sharma2010PRC, JWeiZhou2013PLB, Lenske2015PRC}.

In recent years, the CDF approach with Fock terms, namely the relativistic Hartree-Fock (RHF) theory, was also developed in terms of the density dependent meson-nucleon coupling\cite{PKO12006, PKA12007PRC, Long2010PRC}.
Significant improvements were obtained by the RHF theory, with the involvement of the exchange diagram and the self-consistent tensor force effects \cite{Jiang2015PRC2, Zong2018CPC, WangZh2018PRC}, in describing not only the properties of nuclear ground state\cite{Long2009PLB, Lu2013PRC, Wang2013PRC, LiJJ2014PLB, LiJJ2016PLB2} but the excitation and decay modes\cite{Liang2008PRL, NiuZM2013PLB}.
Besides, several topics on the isospin properties of nuclear matter were studied as well, demonstrating the important role of Fock terms in the nuclear symmetry energy and neutron star properties \cite{Sun2008PRC, Long2012PRC, Jiang2015PRC1, ZQian2015JPG, LiuZW2018PRC}.
As an application of the RHF theory to hot nuclei, the pairing transition in both stable and weakly bound nuclei has already been studied\cite{LJJ2015PRC}. A series of novel phenomenon could occur when the contribution of continuum states is dressed by a finite temperature. For instance, a pairing re-entrance is predicted for drip-line nucleus $^{48}$Ni \cite{LJJ2017PRC}.
However, further systematical study still need to verify the robustness of these predictions. Alternatively, it is interesting to investigate the properties of thermal nuclear matter such as the LG phase transition within these newly developed CDF approaches, especially their model dependence on the selection of effective interactions.

In this work, based on the finite temperature CDF theory with and without Fock terms, the properties of liquid-gas phase transition in thermal symmetric and asymmetric nuclear matter will be studied. The critical parameters of LG phase transition and their correlations with several bulk properties of nuclear matter will be analyzed in detail. The paper is organized as follows. The formalism of the CDF theory for thermal nuclear matter is briefly introduced in Section II. In Sec. III we present the results within CDF calculations and discussion, including the critical point properties of LG phase transition in nuclear matter in Sec. IIIA, the properties of LG phase diagrams in Sec. IIIB, and the correlations between the critical temperature and characteristic pressures in LG phase diagram in Sec. IIIC. Finally, a short summary is given in Sec. IV.

\section{Thermal nuclear matter under the CDF theory}\label{sec:theory}

In this section, the general formalism of the covariant density functional theory for thermal nuclear matter will be described briefly. In order to eliminate the model dependence of the analysis as soon as possible, we will utilize three different meson-exchange types of CDF theory, namely, the relativistic mean-field approach with the nonlinear self-coupling of mesons (denoted as NLRMF), the density-dependent relativistic mean-field (DDRMF) and the relativistic Hartree-Fock (DDRHF) approaches. The corresponding formalism at zero temperature has already been addressed in several references\cite{BanSF2004PRC, Sun2008PRC}.

Based on the meson exchange diagrams of nuclear force, the theoretical starting point --- Lagrangian density can be deduced associated with the degrees of freedom of nucleons ($\psi$), two isoscalar mesons ($\sigma$ and $\omega$), two isovector mesons ($\pi$ and $\rho$), and photons ($A$). For uniform nuclear matter systems, the photon field, describing the electro-magnetic interactions between protons, is ignored naturally. Following the standard procedure \cite{Bouyssy1987PRC}, the energy functional is then obtained by taking the expectation of the Hamiltonian operator $\mathcal H$ with respect to the ground state $|\Phi_0\rangle$,
\begin{align}\label{eq:energy}
E \equiv&\langle \Phi_0|\mathcal H|\Phi_0\rangle= E^{\rm kin} + \sum_\phi \Big( E^D_\phi + E^E_\phi\Big)
\end{align}
where $E^{\rm kin}$ and $E_\phi$ denote the kinetic and potential energy densities, respectively, and for the latter the Hartree-Fock approach leads to two types of contributions: the direct (Hartree) $E_\phi^D$ and exchange (Fock) terms $E_\phi^E$. According to the specific CDF functional $\phi$ could be $\sigma, \omega, \rho, \pi$ etc.. The further details can be found in Ref. \cite{Sun2008PRC}.

The CDF theory at finite temperature is then deduced by considering the grand canonical ensemble in quantum statistical mechanics, where the thermal equilibrium state for a statistical $N$-body system can be determined by the variation of grand canonical potential $\Omega$,
\begin{align}\label{eq:Omega}
  \Omega=&F-\mu N = E-TS-\mu N,
\end{align}
here $F$, $E$, $S$ and $T$ are the free energy, the total energy, the entropy and the temperature, respectively. The associated Lagrange multiplier $\mu$, also referred as chemical potential, is introduced to preserve the particle number at average. Different from the standard CDF approach, the thermal excitation will lead to the spreading of valence particles over the states around the Fermi surface such that the occupation probability $n_i$ of the state $i$ is not 1 or 0 any more.
Therefore, the nucleon density and particle number $N$ read as,
\begin{align}
  \rho_b = & \sum_i n_i u^\dag_i u_i
  , & N = & \sum_i n_i
\end{align}
where $u_i$ is the Dirac spinor for state $i=(p,s,\tau)$, which satisfies the normalization condition $u^\dag_i u_i = 1$. Correspondingly, the entropy $S$ is,
\begin{align}
  S = & -\sum_i \Big[ n_i \ln n_i + (1-n_i) \ln (1-n_i)\Big].
  \label{eq:S}
\end{align}
In this work, the finite-size effects, for instance discussed in Refs. \cite{Qian2001PLB, Piotr2002PRC}, will not be considered for simplicity since the motivation here focuses mainly on systematical exploration of correlations among critical parameters of LG phase transition based on a series of CDF functionals.

The variation of grand canonical potential $\Omega$ shall be performed with respect to the Dirac spinor $u_i$ and the occupation probability $n_i$, respectively, which leads to the nucleon equation at finite temperature, namely the Dirac equation, and Fermi-Dirac distribution $n_\tau(p)$,
\begin{align}
  \big[\bm\gamma\cdot\bm p^* + M^* \big]& u(p,s,\tau) = \gamma_0 \varepsilon^* u(p,s,\tau),\label{eq:Dirac}\\
  n_\tau(p) = & \frac{1}{1 + \exp\big[(\varepsilon (p,\tau) - \mu_\tau)/T\big]},
\end{align}
where $\varepsilon(p,\tau)$ is the single-particle energy of the state $i=(p,s,\tau)$, and the spin index $s$ is omitted since the single particle states are degenerated for $s=\pm1/2$. Notice that the Dirac equation (\ref{eq:Dirac}) is formally unchanged as compared to the one in Ref. \cite{Sun2008PRC}, and the temperature effects lie implicitly in the starred quantities, $\bm p^* = \bm p + \hat{\bm p} \Sigma_V$, $M^* = M + \Sigma_S$, and $\varepsilon^* = \varepsilon(p) - \Sigma_0$,  which are reflected by the occupation probability $n_\tau(p)$ in the self-energies $\Sigma_S$, $\Sigma_V$ and $\Sigma_0$.

After considering the occupation probability induced by the thermal excitation, the energy density functionals, i.e., the kinetic part $E^{\rm kin}$, the potential parts $E_\phi^D$ and $E_\phi^E$ in Eq.~(\ref{eq:energy}) can be obtained as,
\begin{align}
E^{\rm kin}=&\sum_{\tau=n,p}\frac{1}{\pi^2}\int^\infty_0p^2dp (p\hat{P}_\tau+M\hat{M}_\tau)n_\tau(p),\label{eq:EK}\\
E_\sigma^D=&-\frac{1}{2}\frac{g_\sigma^2}{m_\sigma^2}\rho_s^2, ~~
E_\omega^D=\frac{1}{2}\frac{g_\omega^2}{m_\omega^2}\rho_b^2, ~~
E_\rho^D=\frac{1}{2}\frac{g_\rho^2}{m_\rho^2}\rho_{b3}^2,\label{eq:ED}\\
E^E_\phi=&\frac{1}{2}\frac{1}{(2\pi)^4}\sum_{\tau,\tau^\prime}\mathcal I_\phi(\tau,\tau')\int pp^\prime dp dp^\prime n_\tau(p)n_{\tau'}(p^\prime)\nonumber\\
&\hspace{1em}\times \Big[A_\phi + \hat{M}_\tau(p)\hat{M}_{\tau'}(p^\prime)B_\phi + \hat{P}_\tau(p)\hat{P}_{\tau'}(p^\prime)C_\phi\Big],
\label{eq:EE}
\end{align}
where $\mathcal I_\phi(\tau,\tau')$ represents the isospin-related factor, $A_\phi, B_\phi, C_\phi$ are the angle integral coefficients, and $\hat{P}, \hat{M}$ are the hatted quantities,
see Ref.\cite{Sun2008PRC} for details.
The scalar density $\rho_s$, and the baryon density $\rho_b$ and its third component $\rho_{b3}$ read as,
\begin{align}
\rho_s=&\sum_{\tau=n,p}\frac{1}{\pi^2}\int^\infty_0p^2dp\hat{M}_\tau(p)n_\tau(p),\\
\rho_b=&\sum_{\tau=n,p}\frac{1}{\pi^2}\int^\infty_0p^2dpn_\tau(p),\\
\rho_{b3}=&\sum_{\tau=n,p}\tau\frac{1}{\pi^2}\int^\infty_0p^2dpn_\tau(p),
\end{align}
with $\tau=1$ for neutron and $-1$ for proton, respectively.
For NLRMF models, an extra contribution from nonlinear self-coupling of mesons $E_{\rm{N.L.}}$ should be appended in the Hartree terms of potential energies $E_\phi^D$,
\begin{align}
E_{\rm{N.L.}} = -\frac{1}{6}g_2\sigma^3-\frac{1}{4}g_3\sigma^4+\frac{1}{4}c_3\omega_0^4.
\end{align}
After performing the variation to the potential energy densities, the nucleon self-energy $\Sigma(p)$ is obtained, namely
\begin{align}
\Sigma(p)u(p,s,\tau)=\frac{\delta}{\delta \bar{u}(p,s,\tau)}\sum_\phi\left[E_\phi^D +E_\phi^E\right].
\end{align}
Via a self-consistent procedure of self-energies, the properties of thermal nuclear matter can be determined at given density $\rho_b$, the isospin asymmetry $\delta=(N-Z)/(N+Z)$ and the temperature $T$. With the free energy $F$, the pressure of thermal nuclear matter is then derived from the thermodynamic relation,
\begin{align}
P= & \rho_b^2\frac{\partial  }{\partial\rho_b} \frac{F}{\rho_b} =TS+\sum_{i=n,p} \mu_i\rho_{i} -E(\rho_b,\delta,T).
\label{eq:POS}
\end{align}
According to the definition of free energy, the pressure in Eq.~\eqref{eq:POS} can be divided further into
\begin{align}
P=P_{E_0} +P_{E_S} +P_S
\label{eq:P-com}
\end{align}
where the terms $P_{E_0}$ and $P_{E_S}$ are originated from the binding energy per nucleon $E/\rho_b-M$ which is divided further by the isospin symmetric part~$E_0$ and the symmetry energy related one~$\delta^2E_S$, and $P_S$ from the entropy. For instance, the symmetry energy related part $P_{E_S}$ is expressed as
\begin{align}
P_{E_S}=\rho_b^2\frac{\partial}{\partial\rho_b}\left[\delta^2E_S(\rho_b)\right].
\label{eq:PES}
\end{align}
It is clear that the contribution of $P_{E_S}$ is discarded in symmetric nuclear matter as $\delta=0$. For asymmetric matter, $P_{E_S}$ plays a role in the total pressure, and from the definition its value is found to be ascribed qualitatively to the density slope $L$ of symmetry energy at saturation density $\rho_0$. Since $L$ is denoted as
\begin{align}
L=3\rho_b\frac{\partial E_S(\rho_b)}{\partial\rho_b}\Big{|}_{\rho_b=\rho_0},
\end{align}
one then find $P_{E_S}\propto\rho_bL$ approximately at a given density $\rho_b$ (actually fulfilled strictly at $\rho_0$).

To reveal the liquid-gas (LG) phase transition in thermal nuclear matter, one needs to solve the phase coexistence equations,
\begin{subequations}\label{eq:Gibbs}
\begin{align}
  \mu_\tau^L(T,\rho_{b}^L, \delta^L)=&\mu_\tau^G(T,\rho^G_b,\delta^G), \\
  P^L(T,\rho_{b}^L, \delta^L)=&P^G(T,\rho^G_b,\delta^G),
\end{align}
\end{subequations}
which correspond to the Gibbs conditions, i.e., the identical pressures and chemical potentials for both liquid (L) and gas (G) phases at given temperature $T$.
When solving the phase coexistence equations, the stability condition shall be also satisfied as,
\begin{align}
  \rho_b\left(\frac{\partial P}{\partial\rho_b}\right)_{T,\delta}>&0, & \tau\left(\frac{\partial\mu_\tau}{\partial\delta}\right)_{T,P}>&0.
\end{align}

At the critical points of LG phase transition, the temperature, density and pressure of nuclear matter is denoted as $T_C$, $\rho_C$ and $P_C$, respectively.
For symmetric nuclear matter, the critical point is determined by the inflection point of pressure curve with respect to the baryon density, which is,
\begin{align}
  \frac{\partial P }{\partial\rho_b}\Bigg|_{T=T_C} = & \frac{\partial^2P }{\partial\rho_b^2}\Bigg|_{T=T_C} = 0,
  \label{eq:PC1}
\end{align}
while for asymmetric nuclear matter, instead the critical parameters should be solved by the inflection point of chemical potential isobars, namely,
\begin{align}
  \frac{\partial \mu_\tau }{\partial\delta}\Bigg|_{T=T_C} = & \frac{\partial^2\mu_\tau }{\partial\delta^2}\Bigg|_{T=T_C} = 0.
    \label{eq:PC2}
\end{align}

Moreover, one can also introduce the critical incompressibility $K_C$, which is defined as the second derivative of free energy $F$ with respect to the baryon density $\rho_b$ at finite temperature,
\begin{align}
K_C = 9\rho_b^2\frac{\partial^2}{\partial\rho_b^2}\frac{F}{\rho_b}\Big|_{\rho_C}.
\label{eq:K-T}
\end{align}
For symmetric nuclear matter, the first condition at critical point in Eq. \eqref{eq:PC1} can be expressed further as
\begin{align}
\frac{\partial }{\partial\rho_b}\left(\rho_b^2\frac{\partial  }{\partial\rho_b} \frac{F}{\rho_b}  \right)\Bigg|_{\rho_C}=0,
\label{eq:derP}
\end{align}
according to the definition of the pressure Eq. \eqref{eq:POS}. One then readily find a relation between the critical parameters $K_C$ and $P_C$,
\begin{align}
K_C+18\frac{P_C}{\rho_C}=0,
\label{eq:KCPC}
\end{align}
which makes an alternative way to determine the critical point of LG phase transition in symmetric nuclear matter.

\section{Results and discussion}\label{sec:results}
In this work, the analysis based on the CDF theory will be carried out by three kinds of meson-exchange types of CDF functionals: (1) NLRMF functionals NL1\cite{Reinhard1989PPP}, NLZ\cite{NLZNLZ22009}, NLZ2 \cite{NLZNLZ22009}, NL3\cite{NL3NLSH1997}, NL3$^*$ \cite{NL3s2009}, NL-SH\cite{NL3NLSH1997}, NL$\rho$\cite{NLrho2002PRC}, TM1\cite{TM1TM21994}, TM2\cite{TM1TM21994}, TMA\cite{TMA1995}, GL-97\cite{GL972000}, PK1\cite{PKDD2004PRC} and PK1R\cite{PKDD2004PRC}; (2) DDRMF functionals TW99\cite{TW991999}, DD-ME1\cite{DD-ME12002}, DD-ME2\cite{DD-ME22005} and PKDD\cite{PKDD2004PRC}; (3) DDRHF functionals PKO1\cite{PKO12006}, PKO2\cite{Long2008EPL} and PKO3\cite{Long2008EPL}.

In ordinary nuclear matter, the integration over momentum $p$ is carried from zero to Fermi momentum $p_F$. For the nuclear matter at finite temperature, the thermal excitation will lead to the spreading of the valence nucleons over the states nearby the Fermi surface, such that the integration over $p$ shall be done from zero to infinity. Several numerical techniques to this kind of integration have been discussed such as in Ref. \cite{GONG2001CPC}. However, for the cases concerned in this work where the temperature is lower than 20 MeV, the diffusion of Fermi surface is somewhat weak so that the occupation probability $n_i$ drops down to zero promptly. It has been checked that a Gauss-Legendre integration up to about $5p_F$, as momentum cutoff condition adopted here, has guaranteed the convergence numerically in momentum space. Moreover, the phase coexistence equations (\ref{eq:Gibbs}) as a set of non-linear equations are solved numerically with the Powell hybrid method \cite{Powell1970}, which overcomes the deficiency of possible divergence compared to the classical Newton-Raphson method by introducing a "hybrid algorithm" in the iteration of Jacobian matrix.

\subsection{Critical point properties of LG phase transition in thermal nuclear matter}\label{subsec:part1}
Critical parameters are very important characteristic quantities in determining properties of liquid-gas phase transition, among which the critical temperature $T_C$ is especially concerned. For symmetric nuclear matter, $T_C$ is estimated in the range of $10\thicksim20$ MeV in previous studies. To reduce its predicted uncertainty theoretically, the correlations among various critical parameters of LG phase transition account for and need to be discussed not only in symmetric but in asymmetric nuclear matter.

\begin{figure}[t]
\includegraphics[width=0.48\textwidth]{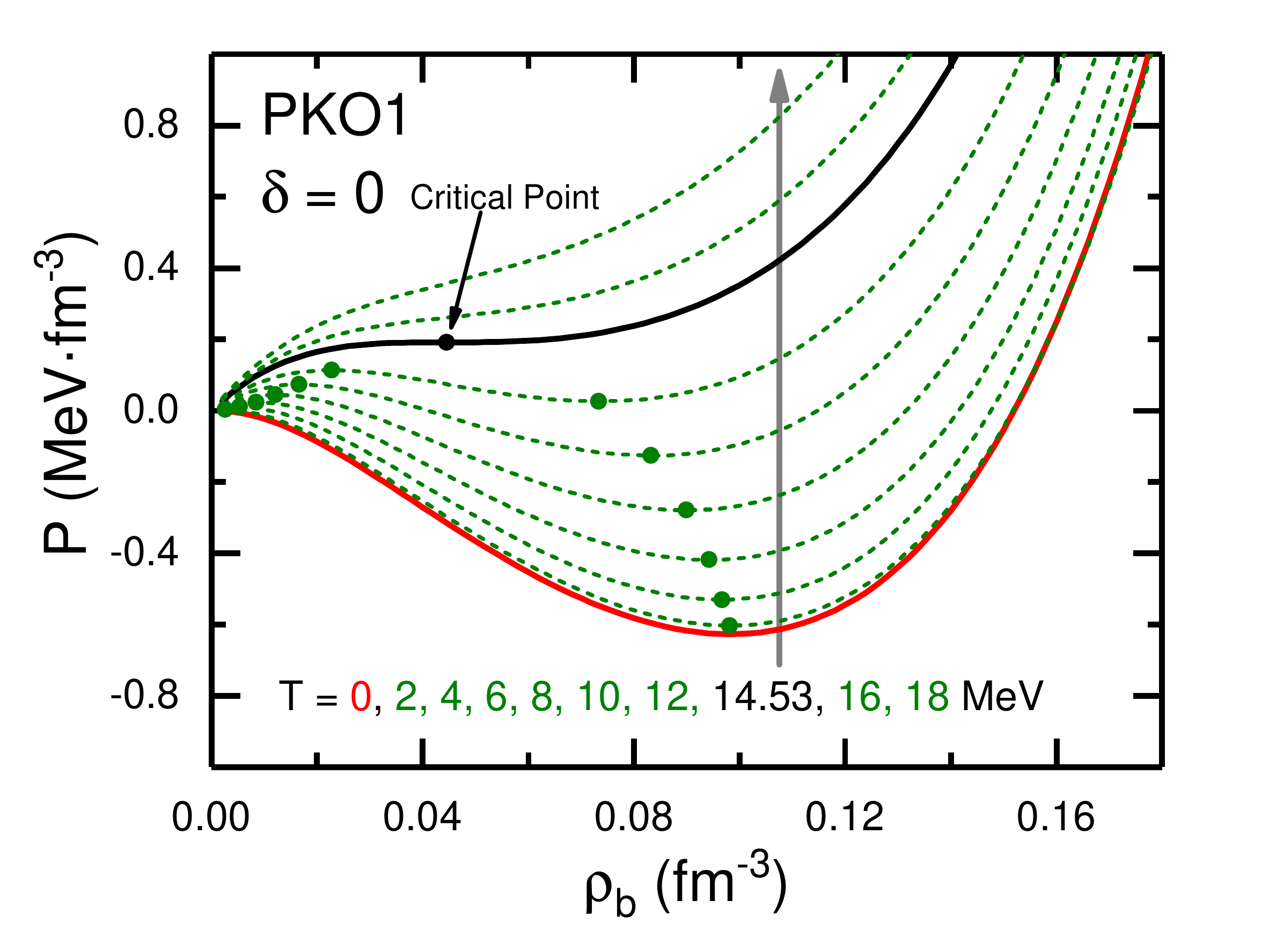}
\caption{(Color Online) Pressure of symmetric nuclear matter ($\delta=0$) as a function of the baryon density $\rho_b$ at various temperatures given by the RHF functional PKO1. In particular, the red (black) solid line represents the result at zero temperature $T=0$ (the critical temperature $T_C$). The filled circles denote the spinodal points (points in which $\partial P/\partial\rho_b=0$) in pressure curves.}
\label{fig:PressTB0}
\end{figure}

For symmetric nuclear matter, the critical point of LG phase transition is determined according to Eq. \eqref{eq:PC1}, which is relevant to the inflection point on $T_C$ isotherm. Taking the RHF functional PKO1 as an example, the calculated pressure curves of thermal symmetric nuclear matter with the baryon density $\rho_b$ are shown in Fig. \ref{fig:PressTB0}. At finite temperature, it is seen that the pressure curves behave a characteristic $S$ shape of van der Waals-like isotherm\cite{Goodman1984PRC, SILVA2008PLB, Rios2008PRC, Rios2010NPA, Vovchenko2015PRC, Lourenco2016PRC, Vovchenko2017PRC}. When the temperature is lower than a certain value which defines the critical temperature $T_C$,
the pressure curve presents a non-monotonic trend with increasing density. Accordingly, the spinodal instability would occur in the density range between two extreme points (points in which $\partial P/\partial\rho_b=0$), leading to the LG phase transition. For PKO1, $T_C$ is found to be $14.53$ MeV, and the critical pressure $P_C$ of LG phase transition is 0.191 MeV$\cdot$fm$^{-3}$ (see Table \ref{Tab:TC-delta0} for others). For classical van der Waals (VDW) gas, it has been deduced that a linear relation between $T_C$ and $P_C$ exists as $T_C/P_C=8b$, where $b$ is the VDW parameter that describes repulsive interaction \cite{Vovchenko2015PRC}. After considering Fermi statistics, the VDW-like equation of state could be established analytically for thermal nuclear matter \cite{Goodman1984PRC, Lourenco2016PRC}, and it is found the $T_C-P_C$ linear relation is still preserved under several approximations such as those to the effective mass and equation of motion of nucleons. In the following, we will check such a relation numerically within CDF functionals.

\begin{figure}[h]
\includegraphics[width=0.48\textwidth]{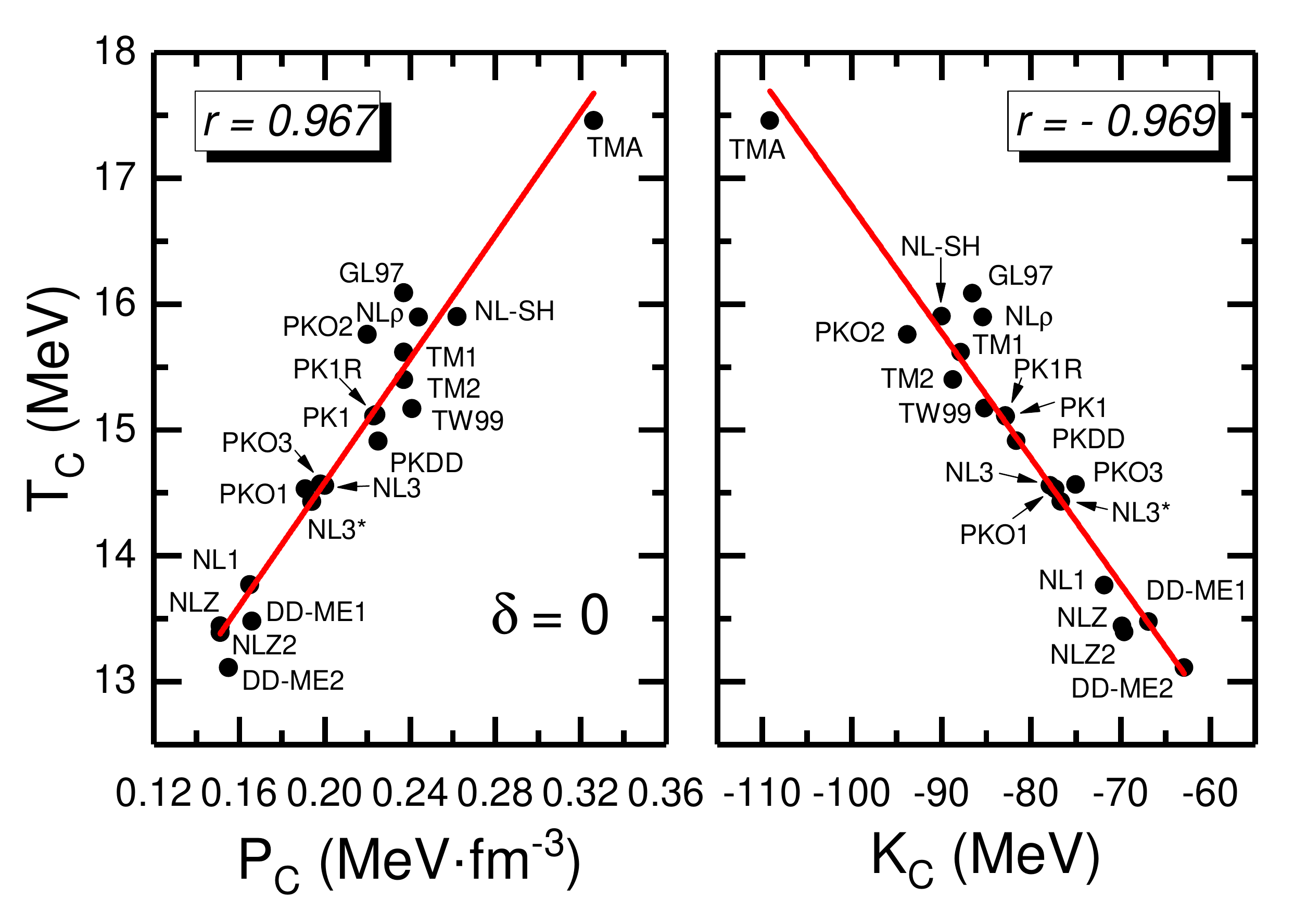}
\caption{(Color Online) For symmetric nuclear matter, the critical temperature $T_C$ of LG phase transition versus the critical pressure $P_C$ (left panel) or critical incompressibility coefficient $K_C$ (right panel). The dots are given by 20 selected CDF functionals and the red lines are from their linear fitting.}
\label{fig:TC-KPB0}
\end{figure}

Figure \ref{fig:TC-KPB0} shows the critical parameters of LG phase transition for symmetric nuclear matter given by the selected CDF functionals, which are determined based on Eq. \eqref{eq:PC1}. It is seen that $T_C$ given by these CDF functionals vary from $13\thicksim 18$ MeV, the range of which is too large to constrain $T_C$ or compare directly with the experimental data. Hence, some model-independent relations or correlations of the critical parameters within themselves or with bulk properties of nuclear matter, once verified, would be very helpful to a better constraint of $T_C$.
In the left panel of Fig. \ref{fig:TC-KPB0}, the critical temperature $T_C$ is plotted versus the corresponding critical pressure $P_C$ given by the selected CDF functionals. A linear correlation between $T_C$ and $P_C$, which has been claimed well in ideal VDW gas, is achieved approximately in present numerical studies.
Then the results can be fitted in terms of
\begin{equation}
T_C=aP_C+b,
\label{eq:TC-b0}
\end{equation}
where $a=24.55~{\rm fm^3}$, $b=9.67~{\rm MeV}$ for symmetric nuclear matter,
and the Pearson's coefficient is $r=0.967$ which indicates notably the robustness of such a $T_C-P_C$ linear correlation to the choice of models.

\begin{table}[t]
\caption{Critical parameters of LG phase transition for symmetric nuclear matter, i.e., the critical temperature $T_C$ (MeV), critical density $\rho_C$ (fm$^{-3}$), critical pressure $P_C$ (MeV$\cdot$fm$^{-3}$),  compressibility factor $Z_C\equiv P_C/(\rho_C T_C)$, and critical incompressibility $K_C$(MeV), as well as the incompressibility $K_0$ (MeV) at saturation density of symmetric nuclear matter at zero temperature.}
\renewcommand{\arraystretch}{1.2}
\begin{ruledtabular}
\begin{tabular}{ccccccc}
            & $T_C$   & $\rho_C$  & $P_C$  & $Z_C$  &
             $K_C$  & $K_0$ \\
            \hline
PKO1      & 14.53  & 0.045 & 0.191  &0.286 & -77.29& 250.24\\
PKO2      & 15.76  & 0.042 & 0.220  &0.332 & -93.83& 249.60\\
PKO3      & 14.57  & 0.048 & 0.198  &0.283 & -75.03& 262.47\\
\hline
PKDD      & 14.91 & 0.049 & 0.225  &0.308 & -81.67& 262.18\\
NL3         & 14.60 & 0.046 & 0.200  &0.297 & -77.90& 271.73\\
PK1         & 15.11 & 0.049 & 0.223  &0.305 & -82.83& 282.68\\
\end{tabular}
\end{ruledtabular}\label{Tab:TC-delta0}
\end{table}

In addition, from the linear relationship between the critical incompressibility $K_C$ and the ratio $P_C/\rho_C$ illustrated in Eq. \eqref{eq:KCPC} for symmetric nuclear matter, it is natural and readily to establish a $T_C-\rho_C K_C$ correlation via $T_C-P_C$ one.
Since the values of $\rho_C$ are close to each other for the CDF functionals, as seen in Table \ref{Tab:TC-delta0}, one would then expect a possible $T_C-K_C$ correlation.
As shown in the right panel of Fig. \ref{fig:TC-KPB0}, the linear correlation between $T_C$ and $K_C$ is verified numerically in CDF approaches, with the Pearson's correlation coefficient $r=-0.969$.
For convenience, one usually introduce a dimensionless parameter to describe such a correlation, namely, the compressibility factor $Z_C$ at critical point of LG phase transition which is defined as
\begin{equation}
Z_C=\frac{P_C}{\rho_C T_C}
\end{equation}
It has been checked that the values of $Z_C$, with samples listed in Table \ref{Tab:TC-delta0} (and also in Table \ref{Tab:TC-delta05} for asymmetric matter), are in general located around $0.3$ from present CDF calculations, in consistence with the previous results analyzed by various density functional approaches \cite{Goodman1984PRC, Rios2010NPA, Lourenco2017PRC}. Furthermore, one notice that these values are also compatible to (although always smaller than) those from standard VDW gas which is known as 3/8 \cite{Vovchenko2015PRC, Lourenco2017PRC}, indicating again the VDW gas-like nature of thermal nuclear matter in CDF approaches.

To better constrain the critical temperature, its dependence on a series of bulk quantities of cold nuclear matter should be investigated as well.
In previous works \cite{Rios2008PRC, Rios2010NPA, Lourenco2016PRC}, it is suggested that the critical temperature $T_C$ of thermal nuclear matter could be correlated with the properties of symmetric nuclear matter at zero temperature such as the incompressibility at saturation density $K_0$. It is essential to confirm the conclusion within various different nuclear models. If a distinct correlation do exists between $T_C$ and $K_0$, the constraint on $K_0$ from a lot of experiments \cite{GARG2018PPNP}, for example the giant monopole resonance \cite{Stone2014PRC}, could be utilized to get more strict value of $T_C$. In Table \ref{Tab:TC-delta0}, in addition to the critical parameters of LG phase transition, we also list the incompressibility coefficients $K_0$ from six characteristic CDF functionals.
However, by checking the Pearson's correlation coefficient (here as $r=0.623$), the quantity $T_C$ is only weakly dependent on $K_0$ in CDF calculations (20 selected functionals). Thus, careful analysis from different parts of the free energy need further so as to find the influence of finite temperature on the incompressibility.

\begin{figure}[t]
\includegraphics[width=0.48\textwidth]{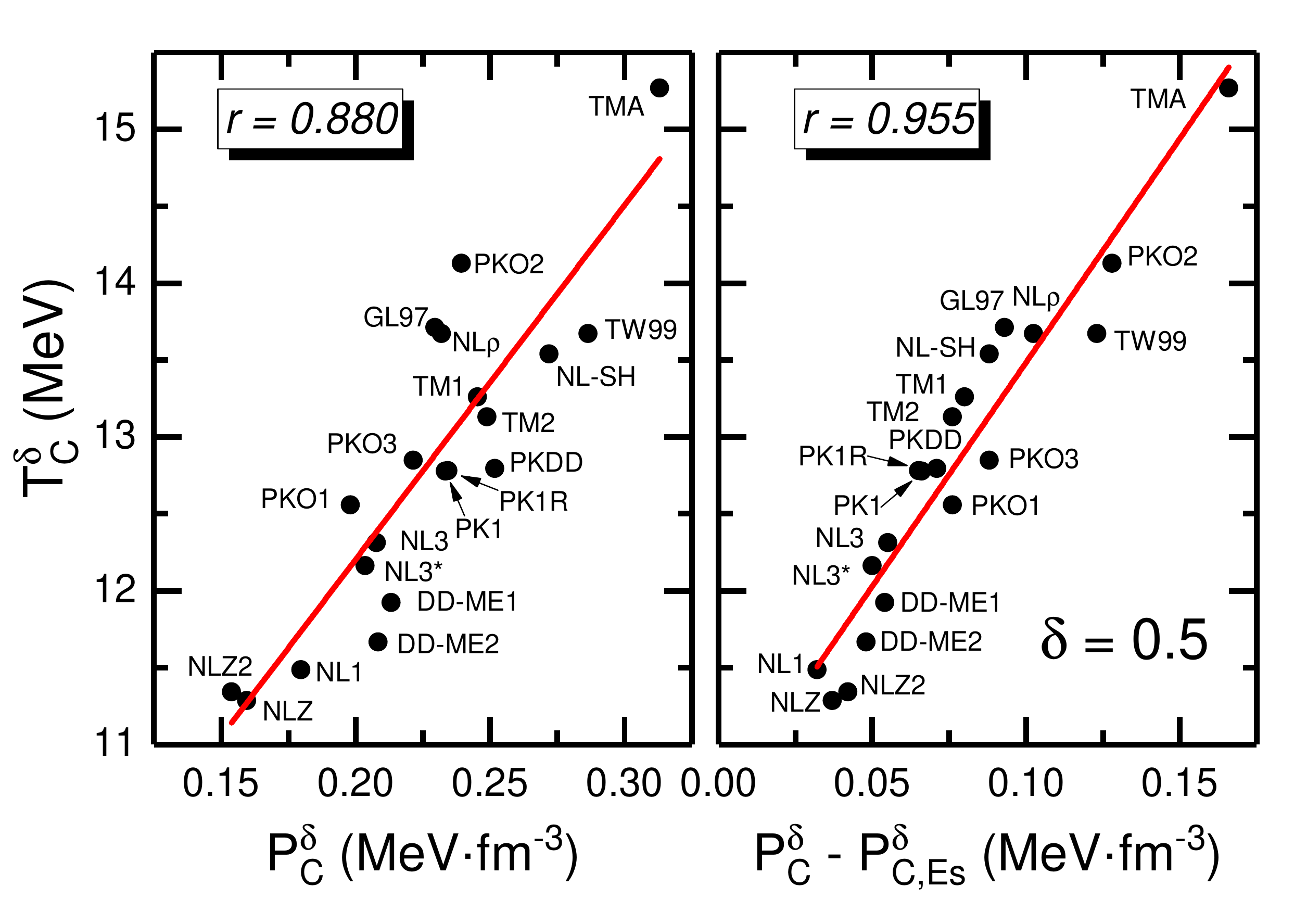}
\caption{(Color Online) For asymmetric nuclear matter with isospin asymmetry $\delta=0.5$, the critical temperature $T_C^{\delta}$ of LG phase transition versus the critical pressure $P_C^\delta$ (left panel) or its contribution with exclusion of the symmetry energy related term $P_C^\delta-P_{C, E_S}^\delta$ (right panel). The dots are given by 20 selected CDF functionals and the red lines are from their linear fitting.}
\label{fig:TC-B05}
\end{figure}

\begin{table}[h]
\caption{Similar as Table \ref{Tab:TC-delta0}, but for the asymmetric nuclear matter at isospin asymmetry $\delta=0.5$.}
\renewcommand{\arraystretch}{1.2}
\begin{ruledtabular}
\begin{tabular}{ccccccc}
            & PKO1  & PKO2  & PKO3 & PKDD & NL3 & PK1 \\
            \hline
$T_C^\delta$      & 12.56 & 14.13 & 12.85 & 12.79 & 12.31 & 12.78 \\
$\rho_C^\delta$ & 0.048 & 0.045 & 0.051 & 0.054 & 0.051 & 0.053 \\
$P_C^\delta$      & 0.198 & 0.239 & 0.221 & 0.252 & 0.208 & 0.233 \\
$Z_C^\delta$      & 0.328 & 0.376 & 0.337 & 0.365 & 0.331 & 0.344 \\
\end{tabular}
\end{ruledtabular}\label{Tab:TC-delta05}
\end{table}

To investigate the change of the feature of LG phase transition with isospin asymmetry $\delta$, it is valuable to look for the possible correlations of LG critical parameters in asymmetric nuclear matter as well. When $\delta\neq0$, the critical point should be determined by the condition in Eq. \eqref{eq:PC2}. The $T_C-P_C$ correlation is checked again but for the case of $\delta=0.5$, as shown in the left panel of Fig. \ref{fig:TC-B05}. In comparison with the symmetric one shown in Fig. \ref{fig:TC-KPB0}, the critical temperature $T_C^{\delta=0.5}$ is no longer linearly correlated well with the critical pressure, as $r=0.880$, while its value locates in the range of $11\thicksim 15.5$ MeV. From Eq. \eqref{eq:P-com}, the contribution to $P_C^\delta$ from different components could be quantified and be helpful to clarify the physical origin of such a destruction of correlation. As compared to the symmetric part $P_{C, E_0}^\delta$ and entropy part $P_{C, S}^\delta$, it is found that the symmetry energy related part $P_{C, E_S}^\delta$ actually has a larger model dependence. Therefore, it is rational that the exclusion of $P_{C, E_S}^\delta$ from $P_C^\delta$, namely $P_C^\delta-P_{C, E_S}^\delta$, exhibits a partly recovered correlation ($r=0.955$) with $T_C^\delta$ for asymmetric nuclear matter, as seen in the right panel of Fig. \ref{fig:TC-B05}.
Therefore, one could introduce a possible linear relation as
\begin{equation}
T_C^\delta=c(P_C^\delta-P_{C, E_S}^\delta)+d,
\label{eq:TC-b05}
\end{equation}
where $c=29.10~{\rm fm^3}$, $d=10.57~{\rm MeV}$ for the case of $\delta=0.5$.
Besides, the Pearson's correlation coefficient between $T_C^{\delta=0.5}$ and $K_{0.1}$ (or $K_0$) is calculated. The smaller value of $r=0.571$ ($r=0.585$) than that in $\delta=0$ case indicates that the correlation between the critical temperature and the incompressibility become worse due to the isospin asymmetry.

\begin{figure}[t]
\includegraphics[width=0.48\textwidth]{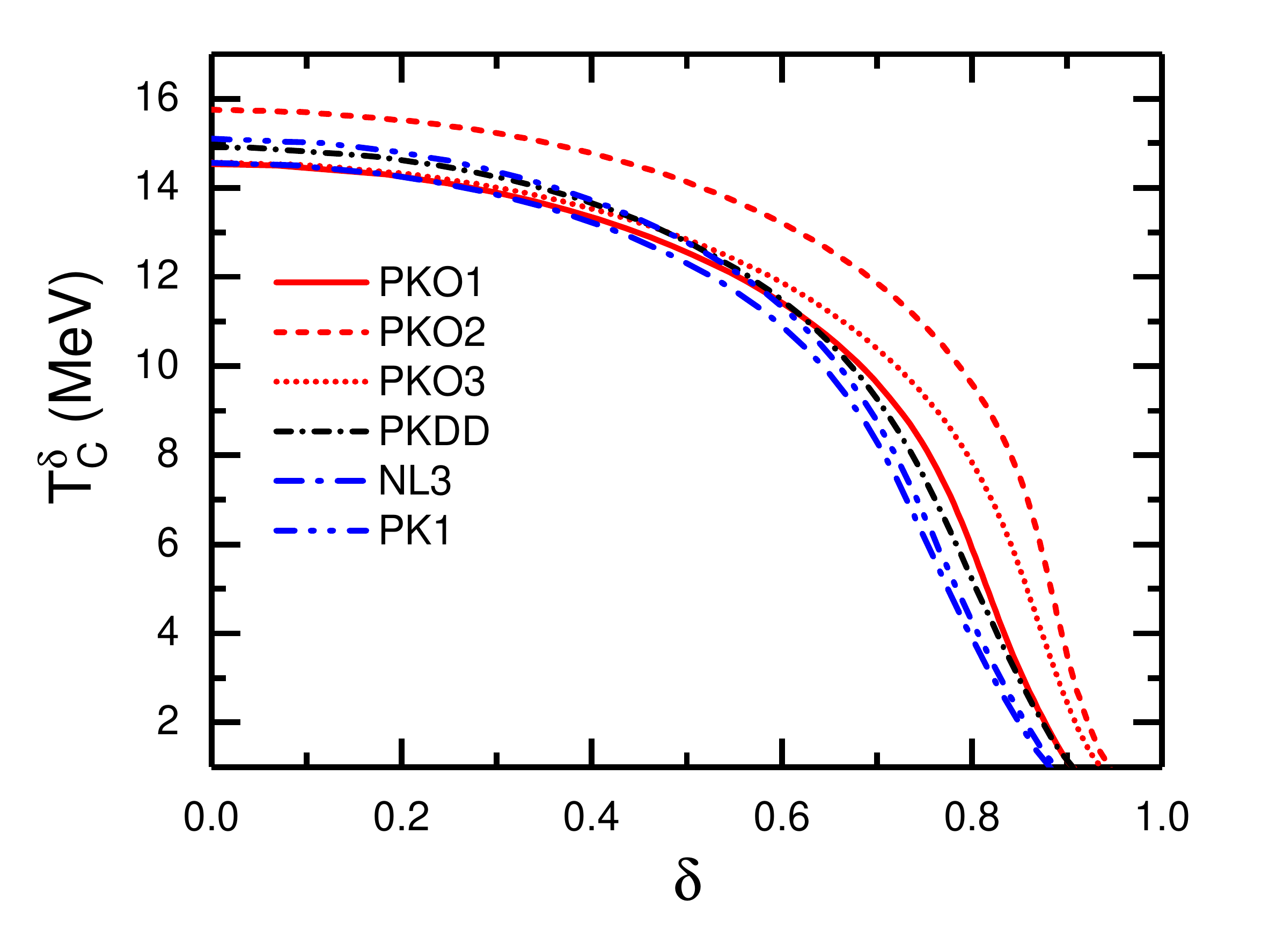}
\caption{(Color Online) The dependence of the critical temperature $T_C^{\delta}$ of LG phase transition on the isospin asymmetry $\delta$ of nuclear matter within selected CDF functionals.}
\label{fig:TC-delta}
\end{figure}

To compare further with the results of symmetric nuclear matter in table \ref{Tab:TC-delta0}, Table \ref{Tab:TC-delta05} shows the critical parameters of LG phase transition at isospin asymmetry $\delta=0.5$.
It is found that the critical temperatures $T_C^{\delta=0.5}$ are smaller than those of symmetric matter, while the critical density and pressure increase slightly as compared to $\delta=0$ case. Recently, one has paid considerable attention to the dependence of LG critical parameters on the isospin asymmetry from experiments of the nuclear caloric curves, where the evolution of the limiting temperature of finite nuclei with mass number and isospin is illustrated \cite{Fevre2005PRL, Sfienti2009PRL, MCINTOSH2013PLB, McIntosh2013PRC}. Here, the isospin dependence of the critical temperature is demonstrated as well within the CDF theory, as seen in Fig. \ref{fig:TC-delta} with several CDF functionals. It is revealed that the critical temperature $T_C^{\delta}$ goes down monotonically with increasing isospin asymmetry $\delta$, in agreement with previous analysis adopting CDF approaches \cite{Sharma2010PRC, JWeiZhou2013PLB}. At small isospin asymmetry, the change of $T_C^{\delta}$ with $\delta$ is moderate, while the value is suppressed drastically after $\delta\gtrsim0.5$, despite a slight model dependence.

\begin{figure}[t]
\includegraphics[width=0.48\textwidth]{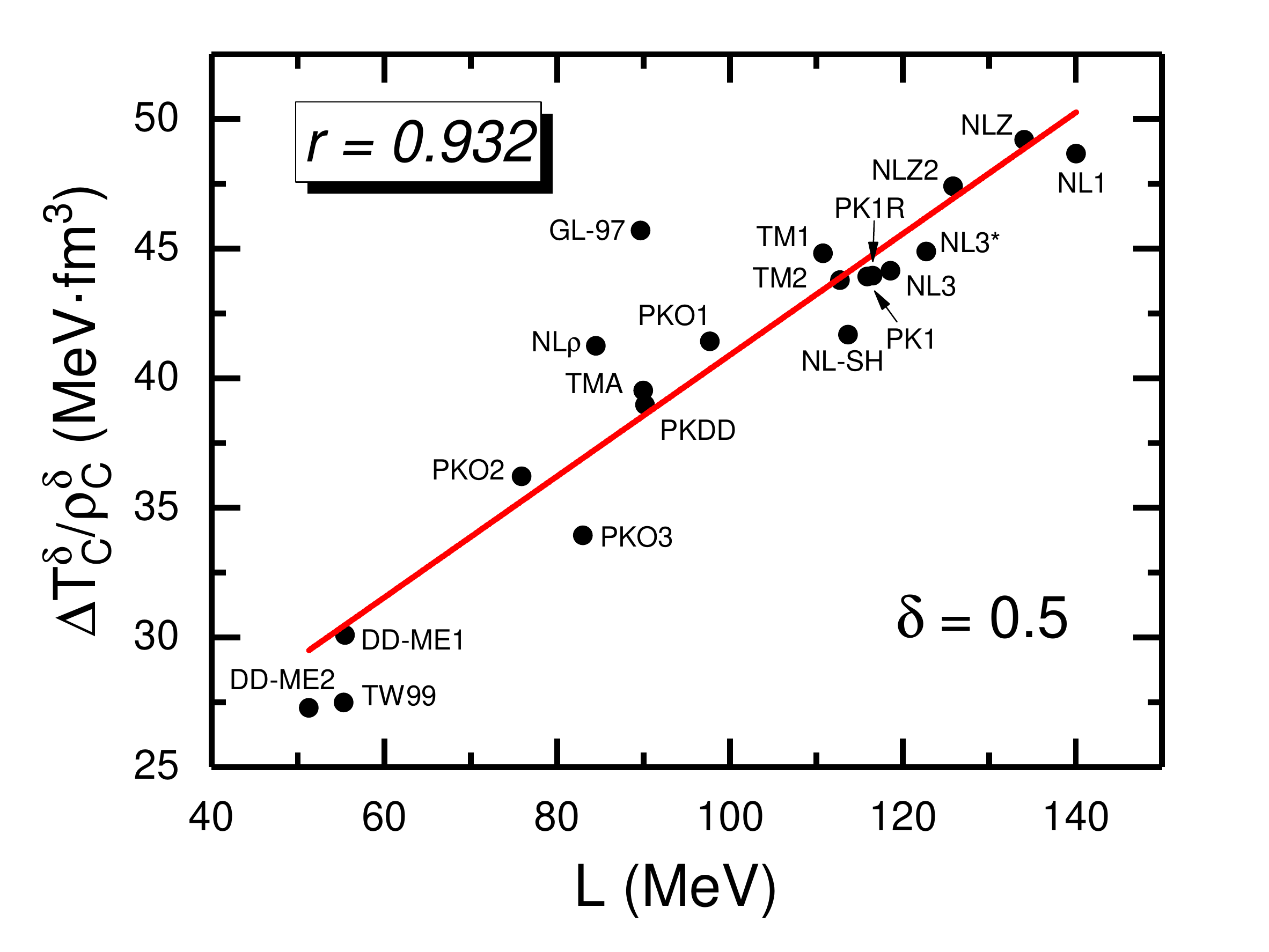}
\caption{(Color Online) The ratio of the scaled critical temperature $\Delta T_C^{\delta}\equiv T_C^{\delta=0}-T_C^{\delta}$ to the critical density $\rho_C^{\delta}$ at isospin asymmetry $\delta=0.5$ versus the density slope of symmetry energy $L$. The dots are given by 20 selected CDF functionals and the red line is from their linear fitting.}
\label{fig:DTCrhoc-L}
\end{figure}

As has been discussed around Eq.\eqref{eq:PES}, $P_{C, E_S}$ could be somewhat related to the symmetry energy, correspondingly being the critical temperature $T_C^{\delta}$ at finite isospin asymmetry. It is helpful to elucidate such a possible relation numerically with present CDF calculations. Assuming the evolution of $T_C^{\delta}$ with $\delta$ is controlled by the symmetry energy (some discussion see Refs. \cite{Sharma2010PRC, JWeiZhou2013PLB}), it is worthwhile to define a scaled critical temperature $\Delta T_C^{\delta}$ for a certain $\delta$ as
\begin{equation}
\Delta T_C^{\delta}\equiv T_C^{\delta=0}-T_C^{\delta}.
\end{equation}
Based on Eqs.~\eqref{eq:TC-b0} and \eqref{eq:TC-b05}, $\Delta T_C^{\delta}$ is then expressed as
\begin{equation}
\Delta T_C^{\delta}= cP_{C, E_S}^\delta+\left[aP_C^{\delta=0}-cP_C^\delta+b-d\right].
\end{equation}
The terms inside square brackets contribute an intercept $\backsimeq[-3.3, -1.6]$~MeV within the selected CDF functionals, showing a weak model dependence.
In combination with the relation $P_{E_S}\propto\rho_bL$, then it is deduced that $\Delta T_C^{\delta}$ is proportional to $\rho_C^{\delta}L$ roughly, as the ratio $\rho_C^{\delta}/\rho_0$ can be treated as an constant approximately in CDF calculations \cite{Lourenco2017PRC}. Utilizing the selected 20 CDF functionals, such a $\Delta T_C^{\delta}-\rho_C^{\delta}L$ relation or $\Delta T_C^{\delta}/\rho_C^{\delta}-L$ relation equivalently is verified numerically, as shown in Fig. \ref{fig:DTCrhoc-L} for the case of $\delta=0.5$. The Pearson's coefficient $r=0.932$ indicates a good linear correlation between the scaled critical temperature and the density slope parameter of symmetry energy. With the constraint on the density slope $L=58.7\pm28.1$ MeV taken from Ref. \cite{Oertel2017RMP}, it is then proposed from Fig. \ref{fig:DTCrhoc-L} that the value of $\Delta T_C^{\delta}/\rho_C^{\delta}$ at $\delta=0.5$ is about $31.2\pm6.6$ MeV$\cdot$fm$^3$.

Hence, it is seen in CDF cases that although the critical temperature of LG phase transition has a clear model dependence, both in symmetric and asymmetric nuclear matter, several linear and model-independent correlations between the critical temperature and other LG critical parameters or bulk properties of nuclear matter could exist. These correlations become worse when nuclear matter is more asymmetric, resulting mainly from the uncertainty of symmetry energy related contributions. However, a linear correlation between the critical temperature at isospin asymmetric case and the density slope of symmetry energy is unveiled, which paves a possible way to constrain the critical parameters of LG phase transition.

\subsection{Properties of LG phase diagram in thermal nuclear matter}\label{subsec:part2}
\begin{figure}[t]
\includegraphics[width=0.48\textwidth]{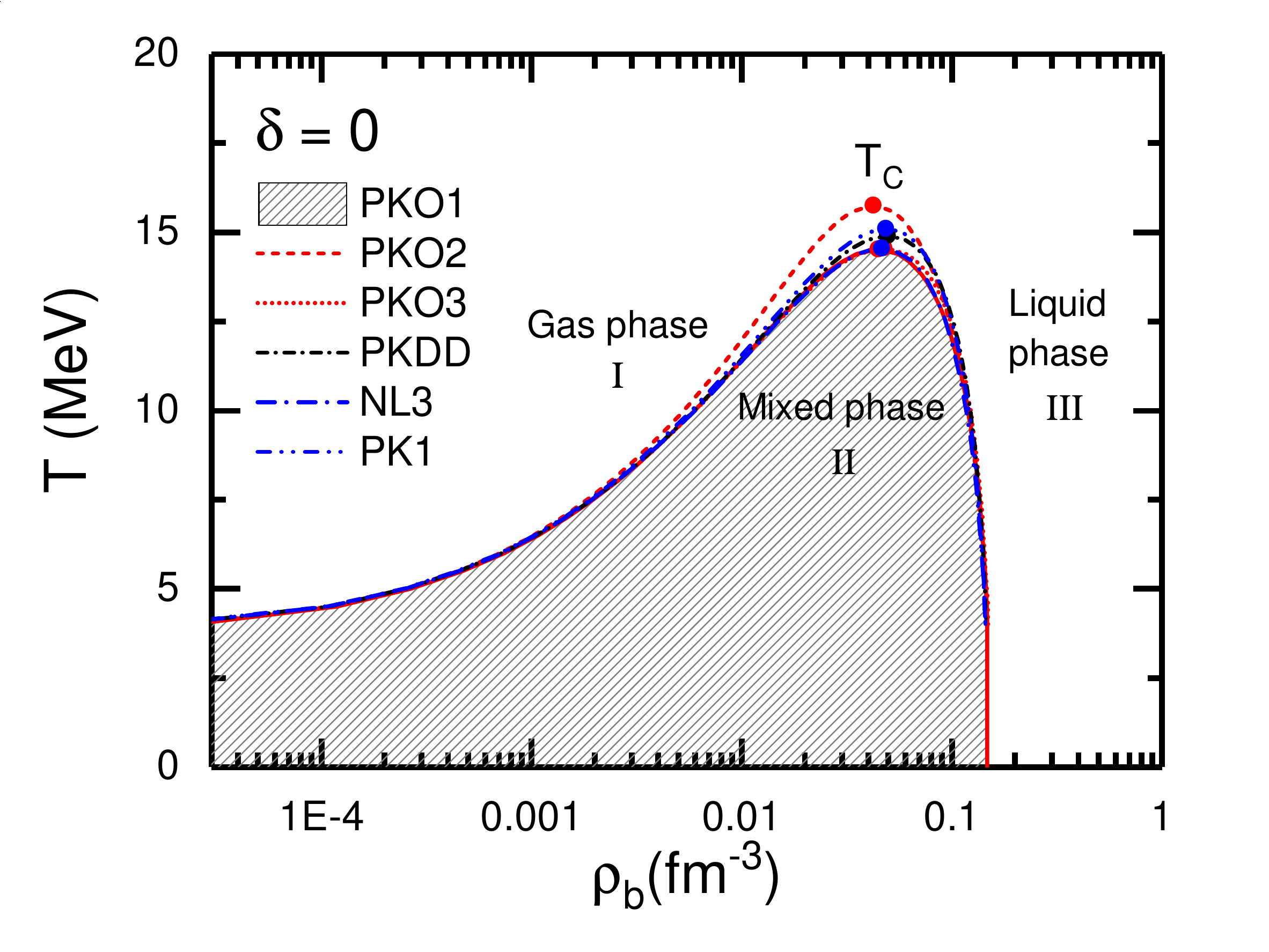}
\caption{(Color Online) Phase diagrams of symmetric nuclear matter at various temperatures as a function of baryon density $\rho_b$(fm$^{-3}$) within CDF functionals, the filled circles denote the critical point of LG phase transition with critical temperature $T_C$.}\label{fig:LGP-T-B0}
\end{figure}

Phase diagram provides essential information about matter structure at a certain circumstance. Specifically, the liquid-gas phase diagram for thermal nuclear matter is substantial to understand several aspects in heavy-ion collision and nuclear astrophysics \cite{Chomaz2004PhysRep, Lattimer2016REP}. Following the above discussion, it is convenient to study LG phase diagrams within CDF functionals, which for the case of symmetric nuclear matter are given in Fig. \ref{fig:LGP-T-B0}. The boundary between two phases can be fixed by solving Eqs. \eqref{eq:Gibbs}. It is found that the phase diagram is divided into three regions: (I) the gas phase at low density; (II) the mixed phase and (III) the liquid phase at high density. Because of the deviation of $T_C$ as listed in Table \ref{Tab:TC-delta0}, there exist an obvious model dependence (particularly the temperature) of the critical points of LG phase transition (filled circles) for the selected CDF functionals, where the RHF functional PKO2 gives the highest $T_C$.

\begin{figure}[t]
\includegraphics[width=0.48\textwidth]{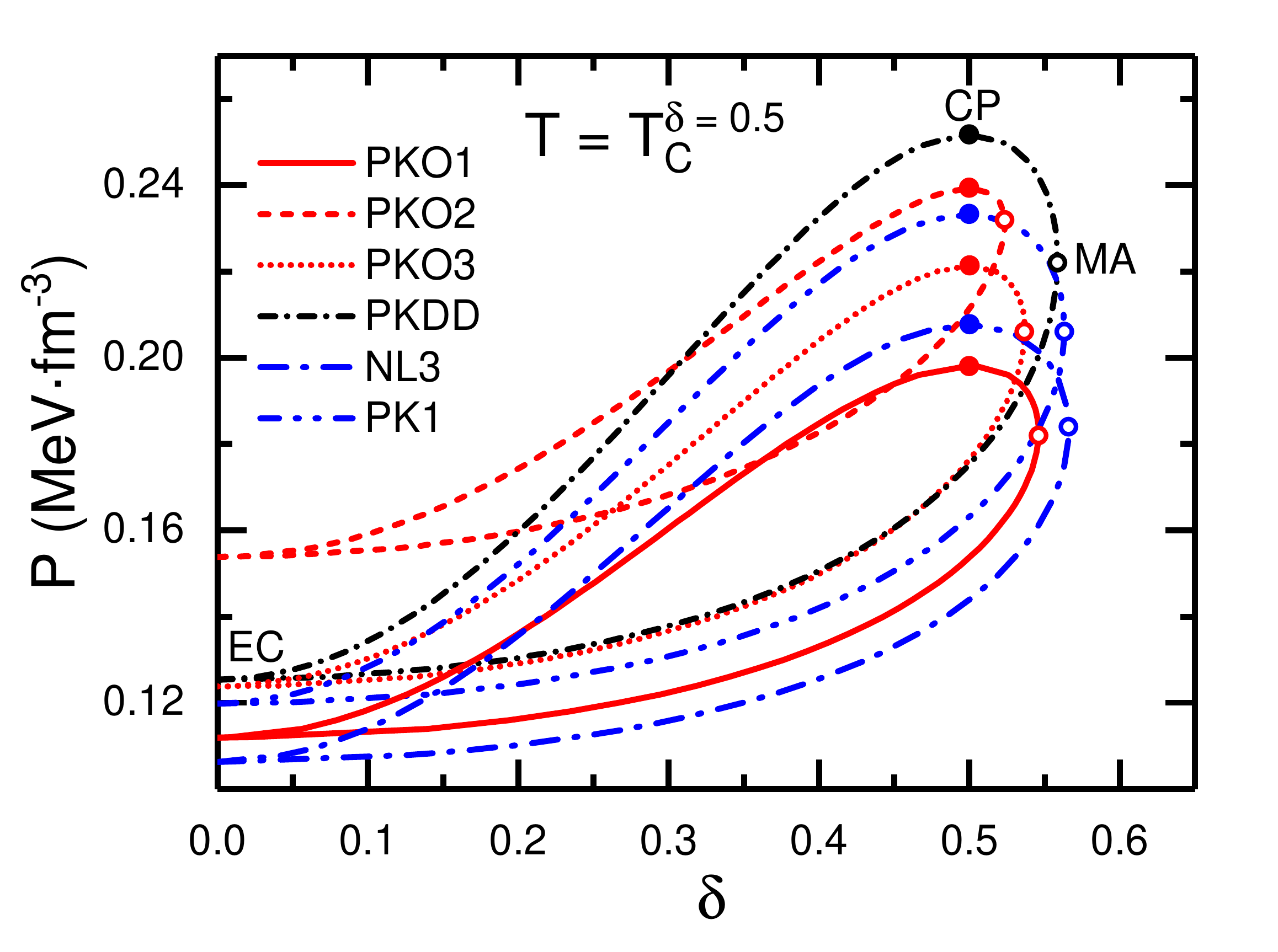}
\caption{(Color Online) Phase diagrams (binodal surface) of thermal nuclear matter at temperature $T=T_C^{\delta=0.5}$, where $T_C^{\delta=0.5}$ for each CDF functional is taken from Table \ref{Tab:TC-delta05}. The filled circles denote the critical pressure (CP) point and the open ones correspond to the maximal asymmetry (MA) point.}\label{fig:LGP-Tc}
\end{figure}

The LG phase diagram for symmetric nuclear matter has also been discussed in many works quantitatively\cite{Rios2008PRC, Rios2010NPA, Lenske2015PRC, Lattimer2016REP}.
For the case of asymmetric nuclear matter, the influence of symmetry energy is supposed to be important in deciding the pattern of the phase diagram \cite{Muller1995PRC, JunXu2007PLB, Sharma2010PRC, JWeiZhou2013PLB}. It is argued that the behavior of liquid-gas phase coexistence could be correlated with the symmetry energy at saturation density \cite{Muller1995PRC} or just its density slope \cite{JunXu2007PLB, Sharma2010PRC, JWeiZhou2013PLB}. To investigate such a topic more quantitatively, here we plot the LG phase diagrams with the selected CDF functionals in Fig. \ref{fig:LGP-Tc} by fixing $T$ for each functional at its own critical temperature $T_C^{\delta=0.5}$ given in Table \ref{Tab:TC-delta05}, which is different from the common treatment of exploring at a constant temperature.

In order to clarify the structure of phase diagram, it is salutary to define three characteristic points: the critical pressure (CP) point (filled circles) determining the maximum pressure $P_{\rm{CP}}$ that the LG phase transition could occur, the maximum asymmetry (MA) point which is given by $\delta=\delta_{\max}$ of the gas phase during the phase transition, and the equal concentration (EC) point of the phase diagram at $\delta=0$. When the phase diagram is plotted in manner of fixing $T=T_C^\delta$, the pressure $P_{\rm{CP}}$ at CP point is just $P_C^\delta$ mentioned in subsection \ref{subsec:part1}.
Correspondingly, the phase diagram is divided into two branches by the CP and EC points, namely the high-density liquid phase line (left branch) and the low-density gas phase line (right branch), and the region surrounded by two lines is the phase coexistence area.
When $\delta$ is larger than one at CP point, namely $\delta>\delta_{\rm{CP}}$ ($\delta_{\rm{CP}}=0.5$ in the case of Fig. \ref{fig:LGP-Tc}), the system will not change completely into the liquid phase \cite{Sharma2010PRC}.
The positions of characteristic points then determine more or less the size of coexistence area, namely, the lower(higher) $P_{\rm{EC}}$ ($P_{\rm{CP}}$) is, the larger the phase coexistence area becomes. Since the small divergence of $\delta_{\max}$ for the selected models as seen in Fig. \ref{fig:LGP-Tc}, one can adopt the pressure difference between CP and EC points, i.e., $P_{\rm{CP}}-P_{\rm{EC}}$, to indicate the size of phase coexistence area of LG phase diagrams. It is seen that $P_{\rm{CP}}$ and $P_{\rm{EC}}$ is clearly model dependent from the picture, leading to the uncertainty of diagram pattern. For instance, a remarkable enhancement of $P_{\rm{EC}}$ is given by PKO2 functional, while its $P_{\rm{CP}}$ is generally comparable with other model predictions, so that a relatively smaller LG phase coexistence area appears in PKO2 case.

\begin{figure}[t]
\includegraphics[width=0.48\textwidth]{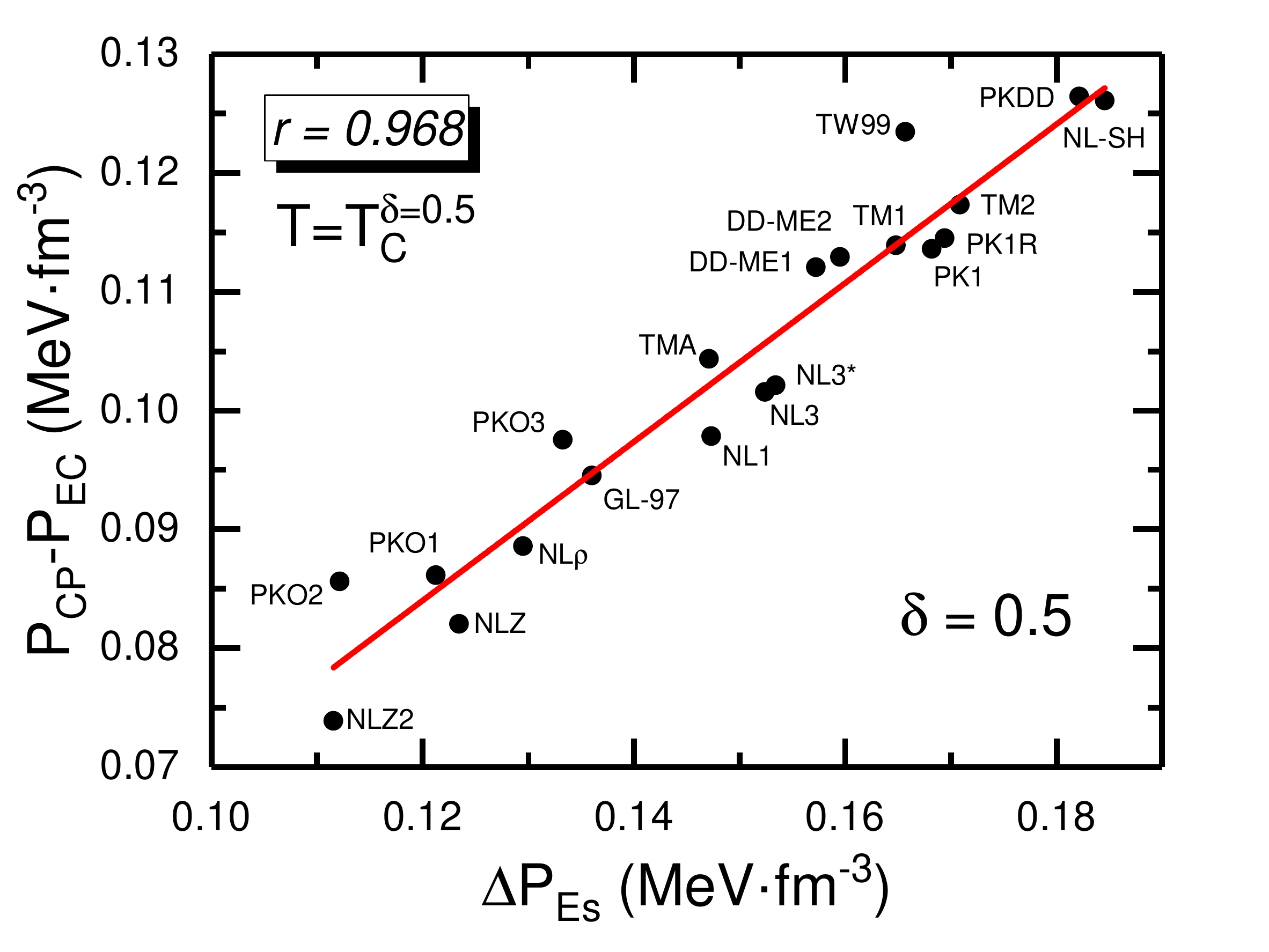}
\caption{(Color Online) For LG phase diagrams at temperature $T=T_C^{\delta=0.5}$, the pressure difference $P_{\rm{CP}}-P_{\rm{EC}}$ versus its symmetry energy part $\Delta P_{E_S}$ taken from Table \ref{Tab:LGP-PC}.}
\label{fig: Pes-DPc}
\end{figure}

From Fig. \ref{fig:LGP-Tc}, it is necessary to extract the pressure values at various characteristic points, as listed in Table \ref{Tab:LGP-PC}, so as to explore how the bulk properties of nuclear matter affect the size of phase coexistence area. For the difference $P_{\rm{CP}}-P_{\rm{EC}}$, PKDD gives the largest value among all functionals, corresponding to the most extensive area of phase coexistence. With the help of Eq. \eqref{eq:P-com}, the contribution of $P_{\rm{CP}}-P_{\rm{EC}}$ can be separated into
\begin{align}
P_{\rm{CP}}-P_{\rm{EC}}=\Delta P_{E_0} + \Delta P_{E_S} + \Delta P_S,
\label{eq:PCPEC}
\end{align}
where $\Delta P_{E_0}, \Delta P_{E_S}$ and $\Delta P_S$ represent isospin symmetric, isospin asymmetric (symmetry energy) and entropy part, respectively. Since $P_{E_S}=0$ at EC point, $\Delta P_{E_S}=P_{C, E_S}^\delta$ which is just the value at CP point. As revealed in Table \ref{Tab:LGP-PC}, the value of $P_{\rm{CP}}-P_{\rm{EC}}$ is mainly ascribed to the contribution of symmetry energy part $P_{C, E_S}^\delta$, while $\Delta P_{E_0}$ and $\Delta P_S$ almost cancel each other although their respective contributions are relatively large. From Eq. \eqref{eq:PCPEC}, one can expect directly a linear correlation between $P_{\rm{CP}}-P_{\rm{EC}}$ and $\Delta P_{E_S}$ as well, which is drawn in Fig. \ref{fig: Pes-DPc}. The Pearson's correlation coefficient is obtained as good as $r = 0.968$, indicating the significant role of the symmetry energy in the size of phase coexistence area in LG phase diagram, in agreement with the conclusion in previous works \cite{Muller1995PRC, JunXu2007PLB, Sharma2010PRC, JWeiZhou2013PLB}. As an alternative case, the linear correlation between $P_{\rm{CP}}-P_{\rm{EC}}$ and $\Delta P_{E_S}$ for LG phase diagram at temperature $T=10$ MeV is done as well, and the above conclusion is confirmed with a correlation coefficient $r=0.898$.

\begin{table}[t]
\caption{The pressure of EC point $P_{\rm{EC}}$ and its difference  $P_{\rm{CP}}-P_{\rm{EC}}$ from the critical pressure for LG phase diagram at temperature $T=T_C^{\delta=0.5}$, taken from Fig. \ref{fig:LGP-Tc}. Correspondingly the components in $P_{\rm{CP}}-P_{\rm{EC}}$ are given according to Eq. \eqref{eq:PCPEC}. The values are in unit of MeV$\cdot$fm$^{-3}$.}
\renewcommand{\arraystretch}{1.2}
\begin{ruledtabular}
\begin{tabular}{ccccccc}
        &  $P_{\rm{EC}}$ & $P_{\rm{CP}}-P_{\mathrm{EC}}$  & $\Delta P_{E_0}$ & $\Delta P_S$  & $\Delta P_{E_0} + \Delta P_S$ & $\Delta P_{E_S}$\\
            \hline
PKO1     &0.112 & 0.086 & -0.442 & 0.407 & -0.035 &  0.121\\
PKO2     &0.154 & 0.085 & -0.403 & 0.377 & -0.026 &  0.113\\
PKO3     &0.124 & 0.097  & -0.469 & 0.433 & -0.036 &  0.135\\
\hline
PKDD    &0.125 & 0.127 & -0.589 & 0.533 & -0.056 &  0.180\\
NL3       &0.106 & 0.102 & -0.530 & 0.480 & -0.050 &  0.150\\
PK1       &0.120 & 0.113 & -0.571 & 0.517 & -0.054 &  0.168\\
\end{tabular}
\end{ruledtabular}\label{Tab:LGP-PC}
\end{table}

\subsection{Correlations of critical parameters in LG phase diagram}\label{subsec:part3}
In subsection \ref{subsec:part1}, the correlations among the critical parameters of LG phase transition in thermal nuclear matter, in particular the critical temperature $T_C$, have been discussed, which is supposed to be a possible way to constrain the critical parameters from bulk properties of nuclear matter. The structure of phase diagram could be reflected in a similar way, if the properties at characteristic points are confirmed to be associated with the critical parameters of LG phase transition as well.

\begin{figure}[t]
\includegraphics[width=0.48\textwidth]{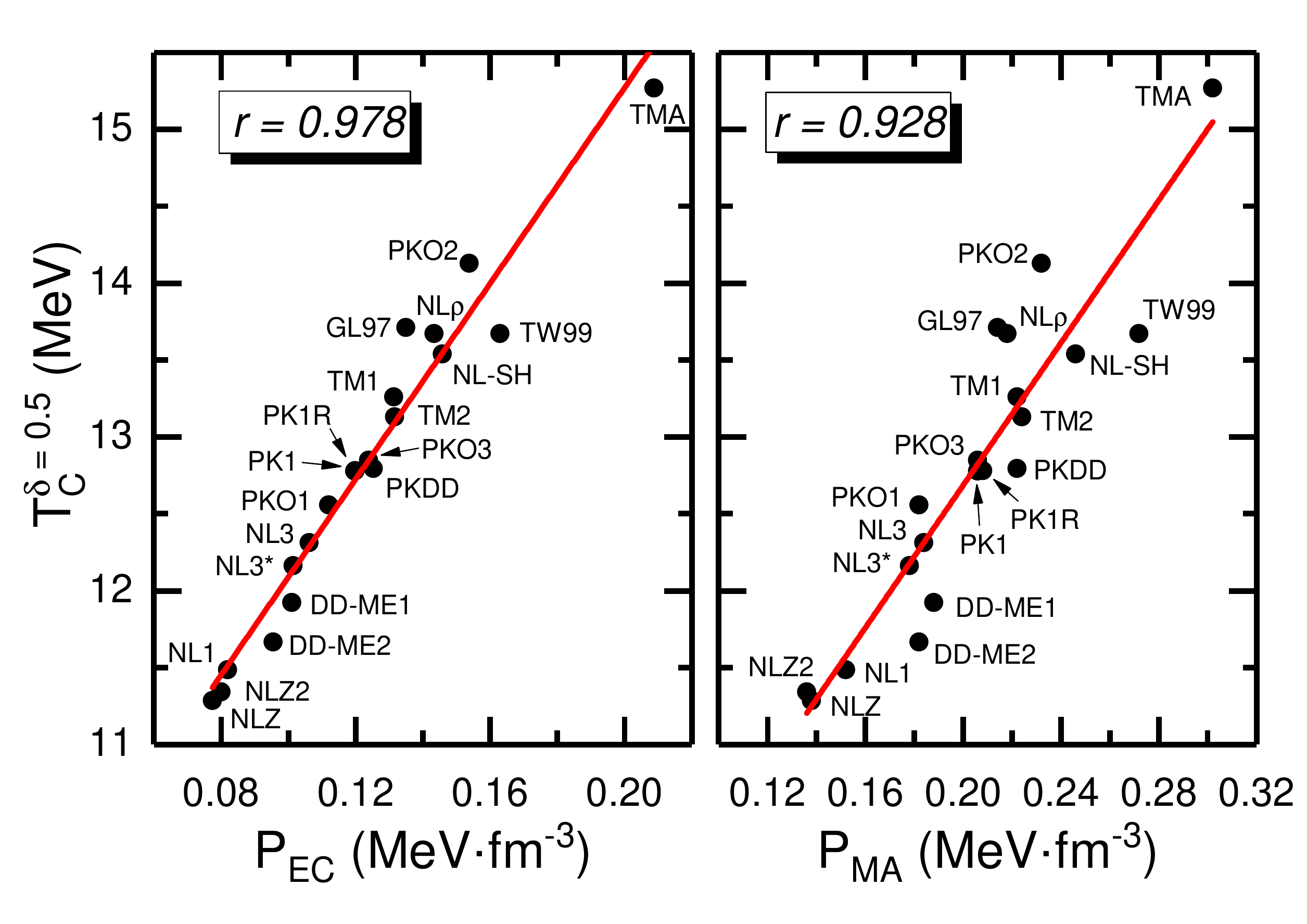}
\caption{(Color Online) For LG phase diagrams at temperature $T=T_C^{\delta=0.5}$ shown in Fig. \ref{fig:LGP-Tc}, the critical temperatures $T_C^{\delta=0.5}$ versus the pressures $P_{\rm{EC}}$ at EC points (left panel) or $P_{\rm{MA}}$ at MA points  (right panel) in phase diagrams. The dots are given by 20 selected CDF functionals and the red lines are from their linear fitting.}\label{fig:TC-PB0M}
\end{figure}

In the left panel of Fig. \ref{fig:TC-PB0M}, the pressures at EC point $P_{\rm{EC}}$ in phase diagrams of thermal nuclear matter at temperature $T=T_C^{\delta=0.5}$ is treated to correlate with the critical parameter $T_C^\delta$ at $\delta=0.5$. Such correlation tend to be well linear with the Pearson's coefficient $r=0.978$, which can be illustrated readily from the satisfied $T_C^{\delta}$ correlation with $P_C^\delta-P_{C, E_S}^\delta$ shown in Fig. \ref{fig:TC-B05} in combination with the relationship deduced from Eq. \eqref{eq:PCPEC} where $P_C^\delta-P_{C, E_S}^\delta$ is equivalent to $P_{\rm{CP}}-\Delta P_{E_S}$. Besides, it is realized from the right panel of Fig. \ref{fig:TC-PB0M} that the pressure at maximum asymmetry (MA) point $P_{\mathrm{MA}}$ is also relevant to $T_C^{\delta=0.5}$ although a relatively smaller $r=0.928$ than one in $P_{\mathrm{EC}}$ case.

\begin{figure}[t]
\includegraphics[width=0.48\textwidth]{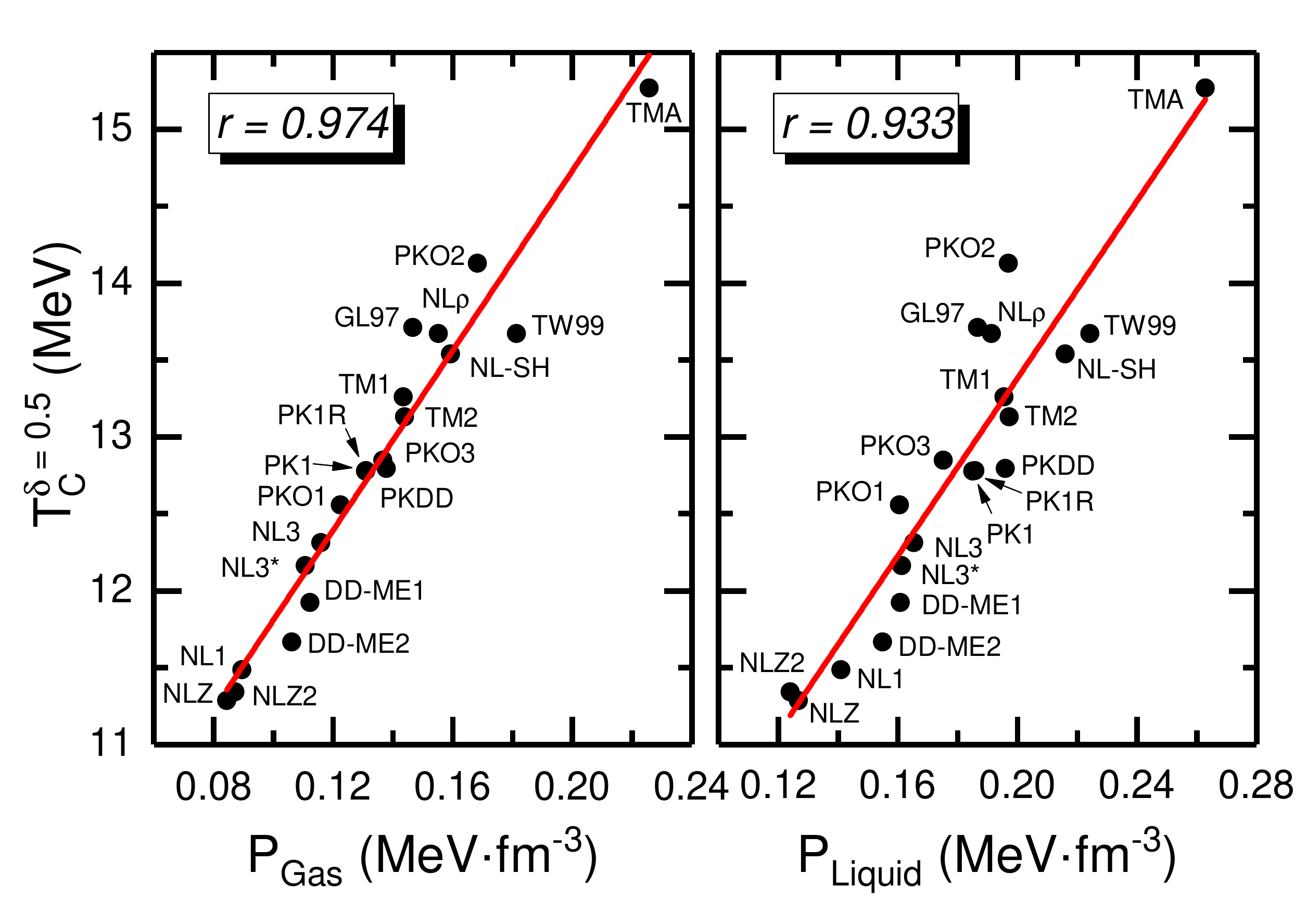}
\caption{(Color Online) For LG phase diagram at temperature $T=T_C^{\delta=0.5}$ shown in Fig. \ref{fig:LGP-Tc}, the critical temperature $T_C^{\delta=0.5}$ versus the pressures $P_{\rm{gas}}$ of gas phase line (left panel) or $P_{\rm{Liquid}}$ of liquid phase line (right panel) which are taken from Fig. \ref{fig:LGP-Tc} with isospin asymmetry $\delta=0.3$. The dots are given by 20 selected CDF functionals and the red lines are from their linear fitting.}\label{fig:TC-PB03}
\end{figure}

At a certain isospin asymmetry $\delta$, two characteristic pressures associated with the gas phase and liquid phase lines can be defined further as $P_{\mathrm{Gas}}$ and $P_{\mathrm{Liquid}}$, which are extracted from Fig. \ref{fig:LGP-Tc}. As plotted in Fig. \ref{fig:TC-PB03} for an example of $\delta=0.3$, both these two quantities are demonstrated to correlate with $T_C^{\delta=0.5}$. Furthermore, it is unveiled that $T_C^{\delta=0.5}$ and $P_{\mathrm{Gas}}$ address a correlation with $r=0.974$, while for $P_{\mathrm{Liquid}}$ case it has a relatively smaller Pearson's coefficient of $r=0.933$, suggesting a better linear correlation for gas phase line than that for liquid phase line. The rule is also proved to be satisfied at other isospin asymmetries, as shown in Table \ref{Tab:r-delta} for $0\leqslant\delta\leqslant\delta_{\rm{CP}}$. It is found that the Pearson's correlation coefficients $r_L$ for the cases of liquid phase line are always smaller than $r_G$ for those of gas phase line when $0<\delta\leqslant\delta_{\rm{CP}}$, which could be interpreted by the fact of a larger CDF model dependence in describing the liquid phase than the gas one since the density of the former is larger and the interaction between nucleons stronger.

\begin{table}[h]
\caption{The Pearson's linear correlation coefficients $r_G$ ($r_L$) between $T_C^{\delta=0.5}$ and $P_{\mathrm{Gas}}$ ($P_{\mathrm{Liquid}}$) at various isospin asymmetry $\delta$ shown in Fig. \ref{fig:LGP-Tc}, see text for details.}
\renewcommand{\arraystretch}{1.2}
\begin{ruledtabular}
\begin{tabular}{cccccccc}
    $\delta$   & $0$   & $0.1$  & $0.2$ & $0.3$ &  $0.4$ & $0.5$  \\
            \hline
$r_G$     &0.978 & 0.978  & 0.977 & 0.974 & 0.969 & 0.957\\
$r_L$     & 0.978& 0.976  & 0.964 & 0.933 & 0.895 & 0.880\\
\end{tabular}
\end{ruledtabular}\label{Tab:r-delta}
\end{table}

From the above discussion, several linear correlations are illustrated between the critical temperature of LG phase transition $T_C^\delta$ at $\delta\neq0$ and the characteristic pressures of LG phase diagrams for asymmetric nuclear matter, including $P_{\rm{EC}}$, $P_{\rm{MA}}$, $P_{\rm{Gas}}$ and $P_{\rm{Liquid}}$, which are demonstrated to be better than $T_C^\delta-P_C^\delta$ correlation as revealed in Fig. \ref{fig:TC-B05} of subsection \ref{subsec:part1}. It is then expected that these correlations could be utilized to constrain the structure of LG phase diagram with the progress in determining $T_C^\delta$ at various isospin asymmetries.

\section{Summary}
In conclusion, by adopting the covariant density functional theory, namely, NLRMF, DDRMF and DDRHF approaches, the liquid-gas phase transition in thermal nuclear matter specifically its properties at critical point has been studied in this work.
The thermal nuclear matter in CDF calculations behaves like van der Waals gas as illustrated in the shape of pressure isotherms and the compressibility factor.
It is seen that the critical parameters, including the critical temperature $T_C$, critical density $\rho_C$, critical pressure $P_C$ and critical incompressibility $K_C$, are clearly model dependent in both symmetric and asymmetric nuclear matter. However, it is verified numerically within CDF functionals that there exist linear correlations approximately between critical parameters and bulk properties of nuclear matter, such as between $T_C$ and $P_C$ ($K_C$).
These correlations become worse for larger isospin asymmetry, which can be attributed from the uncertainty of the contribution $P_{C, E_S}^\delta$ due to the symmetry energy. Correspondingly, the role of the symmetry energy in the isospin dependence of LG transition parameters is focused further. It is unveiled from the CDF calculations that the scaled quantity $\Delta T_C^{\delta}/\rho_C^{\delta}$ can be well determined by the density slope of symmetry energy $L$. Thus, more constraints on nuclear symmetry energy would be crucial and necessary to better understand the critical parameters of LG phase transition at various asymmetric isospin. With recent empirical value of $L$ \cite{Oertel2017RMP}, the value of $\Delta T_C^{\delta}/\rho_C^{\delta}$ at $\delta=0.5$ is suggested to be about $31.2\pm6.6$ MeV$\cdot$fm$^3$.

Then in the later parts of Sec. \ref{sec:results}, the structure of LG phase diagram of thermal nuclear matter is investigated, especially via analysis of the pressure associated with equation of state or entropy. It is found that the size of LG phase coexistence area, determined approximately by the pressure difference $P_{\rm{CP}}-P_{\rm{EC}}$, is well correlated with the pressure part $P_{C, E_S}^\delta$ due to symmetry energy, which is in agreement with the conclusion in previous studies. After extracting the pressure values at several characteristic points in LG phase diagrams, namely, $P_{\rm{EC}}$, $P_{\rm{MA}}$, $P_{\rm{Gas}}$ and $P_{\rm{Liquid}}$, their linear correlations with the critical temperature $T_C^{\delta}$ at non-zero isospin asymmetry are confirmed. Therefore, a possible way is established to depict the pattern of LG phase diagram directly from the critical temperature at virous isospin asymmetries. If $T_C^{\delta}$ can be well constrained such as by the density slope $L$ of symmetry energy, the uncertainty of theoretical prediction to the LG phase diagram will be diminished substantially owing to these correlations, and the physics of liquid-gas phase diagram of thermal nuclear matter will be clarified explicitly.

\begin{acknowledgements}
The authors are grateful to Prof. Wen Hui Long and Dr. Jian Min Dong for helpful discussions. This work is partly supported by the National Natural Science Foundation of China (Grant Nos. 11675065 and 11875152).
\end{acknowledgements}

\bibliographystyle{apsrev}

\end{document}